\documentclass[%
 preprint,
 amsmath,amssymb,
]{revtex4-1}

\usepackage{graphicx}
\usepackage{amssymb}
\usepackage{setspace}
\usepackage[utf8]{inputenc}
\usepackage{array}
\usepackage{natbib}
\usepackage{amsmath}
\usepackage{multirow}
\usepackage{hyperref}
\usepackage{rotating}
\usepackage{adjustbox}

\usepackage{color}
\definecolor{redcolor}{rgb}{1.0,0.,0.}

\definecolor{bluecolor}{rgb}{0,0.,1}

\begin{document}

\preprint{}

\title{Recovering network topology and dynamics via sequence characterization}

\author{Lucas Guerreiro$^1$,  Filipi N. Silva$^2$ and Diego R. Amancio$^1$}

\affiliation{$^1$Institute of Mathematics and Computer Science, University of S\~ao Paulo, S\~ao Carlos, Brazil\\
$^2$Indiana University Network Science Institute, Bloomington, Indiana 47408, USA\\
}


\newpage 

\begin{abstract}
Sequences arise in many real-world scenarios; thus, identifying the mechanisms behind symbol generation is essential to understanding many complex systems. This paper analyzes sequences generated by agents walking on a networked topology. Given that in many real scenarios, the underlying processes generating the sequence is hidden, we investigate whether the reconstruction of the network via the co-occurrence method is useful to recover both the network topology and agent dynamics generating sequences. We found that the characterization of reconstructed networks provides valuable information regarding the process and topology used to create the sequences. In a machine learning approach considering 16 combinations of network topology and agent dynamics as classes, we obtained an accuracy of 87\% with sequences generated with less than 40\% of nodes visited. More extensive sequences turned out to generate improved machine learning models. Our findings suggest that the proposed methodology could be extended to classify sequences and understand the mechanisms behind sequence generation.
\end{abstract}

\maketitle



\section{Introduction}








Many human behavior phenomena are linked to the generation of sequences~\cite{dorle2020learning}. Examples include: the sequence of places people visit in a touristic city~\cite{rodrigues2020tourist}; users visited websites such as social media profiles and posts~\cite{koehn2020predicting}; videos viewed by individuals in a streaming platform~\cite{aguiar2016digital}; everyday decisions, such  as where to eat, where to work; topics of papers produced by scientists, pieces of narratives (in text or movies), or even music~\cite{ferraz2018representation,fell2014lyrics}. In recent years, complex networks~\cite{estrada2012structure} have been used to represent and model the structure of these systems. In most cases, however, the available information is not in the form of networks but sequences.

Some techniques can be used to infer the underlying network from a sequence or set of sequences. Examples are the reconstruction of language models based on text data~\cite{aki2018role,amancio2012using}, where nodes represent words or pieces of text, and human mobility networks~\cite{ramiadantsoa2022existing}, with nodes representing places. In such models, an edge exist for every adjacent nodes appearing in a set of sequences (e.g., of words or of places). Sequences can then be understood as trajectories of nodes that are performed in a network according to a certain dynamics, such as a walk~\cite{ARRUDA2019}. Such a type of reconstruction process is taken place gradually as a discrete knowledge acquisition process~\cite{ARRUDA2017}. To simulate such a process, agents are initially scattered in a network. Next, they perform walks (e.g., random walks) that generate sequences of nodes. For each iteration, each agent partially reconstructs the network from their own history of visited nodes. 

Exploring the properties of the simulated knowledge acquisition process can help understanding real-world problems related to the agents' perception or effectiveness in reconstructing networks from sequences. For instance, in a social media platform, depending on how and how long a user (i.e., an agent) navigates across related posts, they may find a certain post to be central, which may not correspond to the view of other users navigating using different heuristics. In general, the networks reconstructed by users using different walk dynamics can be substantially different. In this context, an important question is if we can recover both the generating dynamics (walks) and the generating network independently. In this work, we explore this question from the perspective of network measurements. In particular, we check if the topological properties (e.g., average degree, transitivity, etc) can be used to recover the network structure and identify the walk dynamics. This differs from previous works~\cite{Guerreiro2020}, in which only the learning curve profiles where used to characterize the generated networks.

Our analysis start with the realizations of well-known random network models, which cover several characteristics present in real-world networks, such as scale-free distribution~\cite{barabasi1999emergence}, small-world~\cite{watts1998strogatz}, presence of community structure~\cite{Lancichinetti2008}, and locality~\cite{Waxman1988}. Then, different walk dynamics are performed in such networks, resulting in sequences of nodes, which are used to create partial reconstructions of each network. Next, network measurements are computed for each network. Finally, we applied classification algorithms to identify the original network model and walk dynamics by looking solely on the generated sequences.

Our results indicate that it is indeed possible to recover the generating model and dynamics from network measurements obtained from sequences. As expected, longer sequences lead to better accuracy. However, we found different combinations of network characteristics and walk dynamics result in different performances for shorter sequences. For instance, the classifiers have difficulty distinguishing from the uniform random models (Erd\H{o}s-Rényi model~\cite{erdos1959}) and geographic models (Waxman model~\cite{Waxman1988}).

The following section discuss related works in the literature as we can see that the concepts of sequences originating from topologies and dynamics are being widely explored and with interesting findings. Later, in the methodology section we present the steps to produce our experiment as well as the configurations of our work. Finally, we present the results and discussions of our study, and the conclusions we have found.

\section{Related Works}


The problem of exploring complex networks via walk dynamics has been addressed in several contexts~\cite{barat1995statistics, meerschaert2006coupled, Comin2020, correa2017patterns}. One particular issue is finding the best dynamics strategy to discover networks nodes (and/or edges) in a optimized way~\cite{lima2018dynamics,ARRUDA2017}. 
In \cite{ARRUDA2017} the authors probed the performance of knowledge acquisition regarding the true self-avoiding dynamics and a Lévy flight-based dynamics on distinct topologies. The influence of variations in the dynamics parameters was analyzed. The authors found that the global impact of parameters variations on knowledge acquisition is surprisingly low. Conversely, the parameter selection is more effective when evaluating the knowledge acquisition problem locally. All in all, this study observed that, when performing collective discovery in a structure, the dynamics parameters do not affect significantly the performance. 


In a related study,~\cite{Guerreiro2020} investigated how knowledge is acquired on different configurations of models and dynamics. The learning time regarding nodes discovery rate in a given period was the main focus of the study. The authors reported that the efficiency in acquiring knowledge depends on how the network is explored and on the topology of knowledge representation. 
The true self-avoiding dynamics was found to be effective in most scenarios. Most importantly, the same learning behavior could be achieved with different pairs of network structures and agent dynamics.  This means that it is possible to generate the same efficiency in learning by changing both knowledge organization and the way knowledge is acquired.


The problem of analyzing how agents learn the structure of networks has also been explored with variation in the way knowledge is transmitted and stored. In \cite{lima2018dynamics} the concept of a \emph{network brain} was proposed. The idea behind this model comprehends a centralizing structure that receives the knowledge discovered by multiple agents walking over the network. Interestingly, the authors report that neither the topology nor the dynamics strongly impacts the learning efficient. 

Some works also have investigated random walks as generator of symbols  data~\cite{ARRUDA2019,Guerreiro2020}. In~\cite{ARRUDA2019} the authors argue that a sequence of symbols can be seen as being generated from walks on a networked topology. The authors analyzed how the performance of different combinations of topologies and dynamics may impact the performance of knowledge acquisition. The study also analyzed how well the network is reconstructed when it is transmitted as a sequence of visited nodes that are compressed before transmission. The so-called knitted networks 
-- similar to word adjacency networks -- were found to display the best performance when considering both compression efficiency and recovery of sequential data. 


The reconstruction of networks is another relevant topic related to the current study~\cite{Lacasa2008, Zhang2006, Gao2009, Fang2018}. One of the most well-known methods is the visibility graph~\cite{Lacasa2008}.  The method is based on the idea that each node in a time series of values can see other nodes, and the spatial visibility criteria is used to generate edges. Such rather simple idea has been proven to be effective on the reconstruction of networks, being able to translate, for example, from a periodic time series to a regular graph, while a a random series as a random graph. While the work proposes a reconstruction method, it does not consider that the original time series stems from a particular walk dynamics. 

Differently from previous works, here we aim at identifying the main properties of the reconstructed networks. More specifically, we investigate whether the reconstruction of  networks via co-occurrence strategy is able to identify the underlying topology and dynamics generating the observed sequence of symbols. 


\section{Methodology} \label{sec:methodology}

\begin{figure}[h]
\centering\includegraphics[width=1\linewidth]{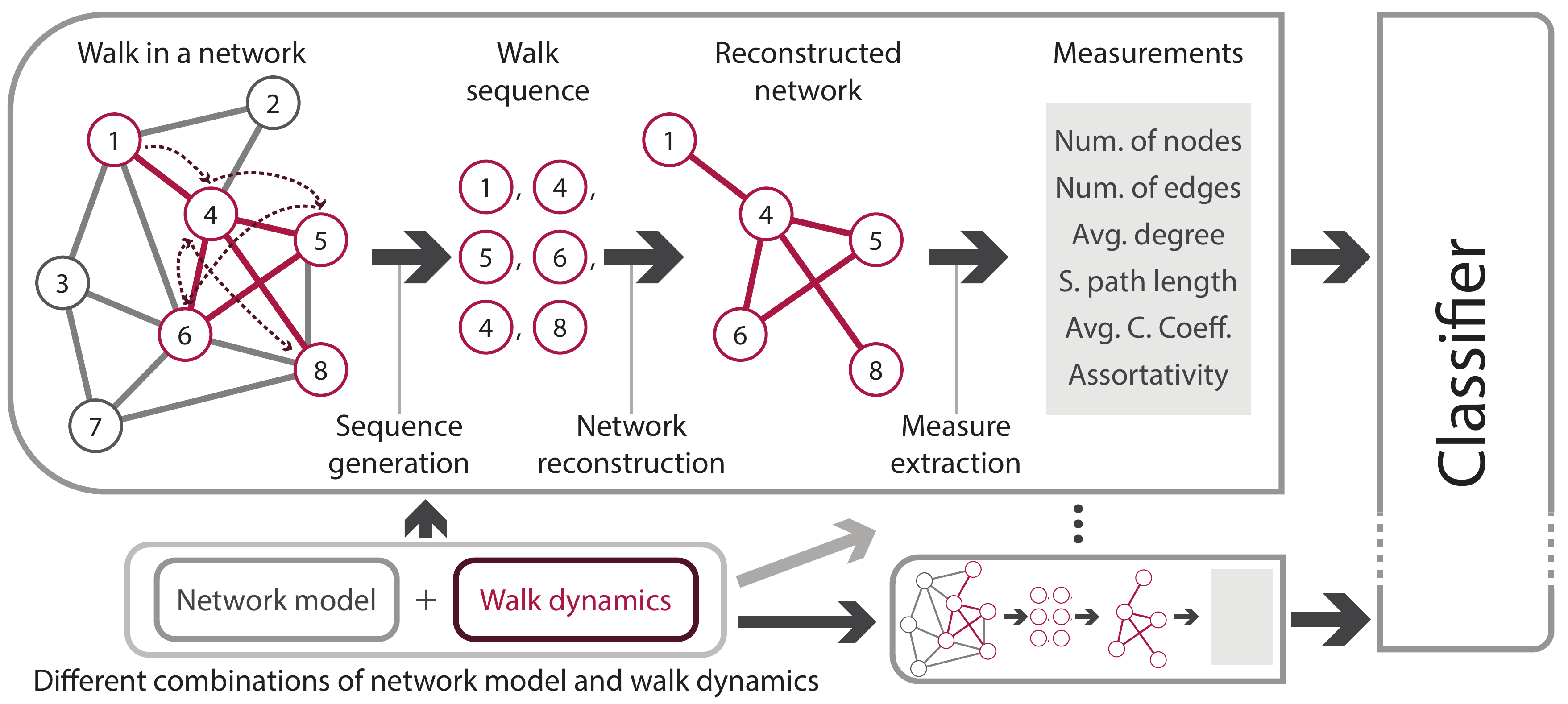}
\caption{Schematics of the methodology. First, we generate multiple iterations of the considered network models. Then, for each considered walk dynamics, we sample a number of walk sequences. Each walk sequence is then used to generate a reconstructed network. This is accomplished by connecting nodes according to adjacency of nodes in the sequence. For instance, for a sequence of nodes $\{1,4,5,6,4,8\}$, edges $\{(1,4),(4,5),(5,6),(4,8)\}$ are reconstructed. Next, we compute a set of network measurements for each walk sequence. Finally, the features of each network are used to train a classifier that try to recover the corresponding model, walk dynamics or both.}
\label{fig:schematic}
\end{figure}

In this paper, we probe whether the characteristics of time series can be useful to infer the topology and walk dynamics associated to the observed sequence, i.e. the sequence generated by walking over the network. We adopted the following methodology to test our hypothesis. First, we generated artificial network topologies. Well-known random walks were also used to generate the sequence of symbols. The obtained sequences are used to create a network, in the network reconstruction step. Then, the structural properties of the reconstructed are extracted, and they are used as input to the machine learning method aiming at identifying the associated topological and dynamical properties used to generate the sequence. 
The proposed framework is illustrated in Figure \ref{fig:schematic} and the main steps are summarized below.

\begin{enumerate}

    \item \emph{Network topology}: we selected four well-known network models to generate the topology of the networks. The chosen topologies includes random networks and models with more realistic features, including heterogeneity in connectivity.
    
    \item \emph{Network dynamics}: given a network, an agent walks over the network in order to generate a sequence of symbols. We also have used four well-known network dynamics to explore network. As a result, a sequence of visited nodes is generated. 
    
    
    \item \emph{Network reconstruction}: the sequence of generated symbols by the agent walking through the network is used to reconstruct the network. This process uses a co-occurrence approach  to create a representation of the original network. 
    
    
    \item \emph{Properties extraction}: features are extracted from the reconstructed networks. The extracted features are then used in the machine learning algorithms. We used 6 features to characterize the structure of the reconstructed networks.
    
    
    \item \emph{Structures classification}: features are fed into different machine learning methods in order to recover both topology and the dynamics that generated the observed sequence. In summary, given an unknown sequence, we aim at identifying the network model and dynamics that generated it. 

\end{enumerate}

\subsection{Network topologies} \label{sec:topologies}

We considered four well-known network topologies. The adopted network parameters are similar to those employed in similar papers studying the relationship between network topology and dynamics~\cite{Guerreiro2020}. 
%
%
We adopted the following undirected and unweighted network models~\cite{erdos1959, erdos1960,Lancichinetti2008}:

\begin{itemize} 

    \item \emph{Erd\H{o}s-Rényi (ER)}: this model generates random graphs. To create this network, each candidate edge is established based on a global probability ${p}$. Random networks are typically characterized by low shortest path lengths and low values of local connectivity (clustering coefficient). 
    
    \item \emph{Barabási-Albert (BA)}: the BA model~\cite{barabasi1999emergence} reproduces the scale-free distribution of node degree. The model adds, at each step, a new node the has probability of linking to other nodes of the network. Let $k_i$ be the degree of node $i$. The probability of $i$ to receive a link from the new node ($p_i$) is proportional to $k_i$, i.e. $p = k_i / \sum_j k_j$. Because a new node has a preference to connect with more connected nodes, a few hubs arise.
    
    \item \emph{Waxman (Wax)}: this model implements a geographic network. In order to construct such a network, all nodes are randomly distributed into a two dimensional space. The probability of a link existing between two nodes takes into consideration the distance between nodes, so that nearby nodes have a higher probability of being connected to each other \cite{Waxman1988}. 
    
    
    \item \emph{Modular Networks (LFR)}: we have also used a topology that reproduces the modularity of real-world networks. We adopted the implementation proposed in~\cite{Lancichinetti2008}.  This model also reproduces the scale-free behavior within network community. The parameters employed to generate the communities is similar to those used in related studies~\cite{Guerreiro2020}. The following parameters were considered here: number of communities ($n_C$), exponent for the degree sequence $(t_1)$, community size distribution ($t_2$) and mixing parameter ($\mu$)~\cite{Lancichinetti2008}. As in related works, we adopted the following values: $n_C=5$, $t_1=3$, $t_2 = 0$ and $\mu=0.2$. 
    The maximum node degree is chosen in order to obtain the desired average degree of the network.
    
    
\end{itemize}

We considered networks comprising $N = 5000$ nodes with the following values of average degree $\langle k \rangle = \{4,6,8,10\}$. A wider range values for $N$ was not considered because, in preliminary experiments, we have not observed significant variations in the obtained results.

\subsection{Network dynamics} \label{sec:dynamics}

The networks are explored with four different agent dynamics. We have selected the traditional random walk dynamics (RW)~\cite{Lovasz1996} and three variations of this random walk: degree-biased random walk (RWD)~\cite{Bonaventura2014}, random walk biased towards the inverse of the degree (RWID)~\cite{Bonaventura2014}, and the true self-avoiding walk (TSAW)~\cite{Kim2016,Amit1983}. As illustrated in Figure \ref{fig:schematic}, the agent dynamics are the rules to walk over the network. The adopted dynamics are described below:

\begin{itemize}

	\item \emph{Random Walk (RW)}: the traditional random walk randomly chooses one of the neighbors of the current node based on a uniform distribution. The probability of transition from node $i$ to $j$ is $p_{ij} = k_{i}^{-1}$, where  $k_i$ is the degree of $i$.

	\item \emph{Random Walk biased towards the Degree (RWD)}: this dynamics is a variation of the traditional RW. Here, the agent has a higher probability of visiting nodes with larger degree. The transition probability $p_{ij}$ is computed as:
    \begin{equation}
        p_{ij} = \frac{k_{j}}{\sum_{l \in \Gamma_{i}} k_{l}},
    \end{equation}
    where $\Gamma_{i}$ is the set comprising the neighbors of $i$.

	\item \emph{Random Walk biased towards the Inverse of the Degree (RWID)}: here the agent tends to visit nodes with smaller degrees, according to the following equation for the transition probability:
    \begin{equation}
        p_{ij} = \frac{k_{j}^{-1}}{\sum_{l \in \Gamma_{i}} k_{l}^{-1}}.
    \label{eq:rwid}
    \end{equation}
	\item \emph{True Self-Avoiding Walk (TSAW)}: this dynamics is a variation of the self-avoiding random walk~\cite{Herrero2005saw}. In this dynamics, the agent avoids visiting nodes previously visited. As observed in related works, this walks tends to explore more efficiently the network~\cite{ARRUDA2019, Guerreiro2020}. The probability transition for the TSAW is computed as:
    \begin{equation}
        p_{ij} = \frac{ e^{-\lambda f_{ij}}}{\sum_{l\, \in\, \Gamma_{i}} e^{-\lambda f_{il}}},
    \label{eq:tsaw}
    \end{equation}
    where $f_{ij}$ denotes the frequency that the edge linking nodes $i$ and $j$ has been visited. The parameter $\lambda$ is a positive constant that controls how likely an agent will visit an edge that has already been visited. Similar to related studies, we are using $\lambda = \ln 2$~\cite{ARRUDA2019, Guerreiro2020}.

\end{itemize}

In order to analyze the dependence of the results with the length of the random walk ($S$), we considered different values of $S$. This includes very short walks and also a number of steps compatible with the network size considered. We used $(S) = \{10, 50, 100, 500, 1000, 2000, 5000\}$.

\subsection{Network reconstruction} \label{sec:reconstruction}

The next step in the framework presented in Figure \ref{fig:schematic} is to reconstruct the network. The network is reconstructed based on the resulting time series of visited nodes. Each pair of adjacent nodes observed in the sequence of visited nodes is linked with an edge. This is similar to the co-occurrence strategy used to create networks from sequential data. This co-occurrence model is particularly used when modeling texts as networks~\cite{aki2018role,amancio2012using,machicao2018authorship,stella2021mapping,liu2013language}. For each network model and random walk, we considered 1000 realizations. 


\subsection{Properties extraction} \label{sec:properties}

We have selected the following properties to characterize the reconstructed networks:

\begin{enumerate}

    \item \emph{Number of nodes}: the size of the reconstructed network will also represent the number of nodes discovered after the random walk. We intend to analyze whether the total of discovered nodes is a relevant information to identify the topology and dynamics used to generate the sequence. Related works have shown that the number of discovered nodes can vary according to the adopted topology and dynamics~\cite{Guerreiro2020}, with the highest efficiency observed for the   TSAW random walk. This measure alone, however, is not enough to discriminate all combination of network topology and agent dynamics~\cite{Guerreiro2020}.
    
    \item \emph{Number of edges}: similarly to previous property, we extracted the number of discovered edges, i.e. the total number of edges of the reconstructed network. 
    
    \item \emph{Average degree}: we also explore the influence of the reconstructed networks’ average degree on the identification of the originating structures. Although this information can be recovered from the both number of nodes and edges, we included this information in order to discuss the results in terms of this well-known quantity as well.
    
    \item \emph{Shortest path length}: this property quantifies the typical shortest path length between all nodes of the reconstructed network. We also measure -- in the reconstructed network -- the shortest path length between the first and last nodes of the sequence used to reconstruct the network.
    
    \item \emph{Average clustering coefficient}: this property quantifies the local density of edges of a node, which is directly related to the number of triangles. The clustering coefficient $C_i$ of a given node $i$ is:
    \begin{equation}
        C_i = \frac{2 T(i)}{k_i (k_i - 1)},
        \label{eq:clustering}
    \end{equation}
    where $T(i)$ denotes the number of triangles including node $i$. Equivalently, $T(i)$ is the number of links between neighbors of $i$. 
    
    \item \emph{Assortativity}: this coefficient measures whether nodes with high degree tends to be linked with other highly connected nodes. The assortativity can be measured in the of the degree correlation of linked nodes~\cite{yuan2021assortativity}.  
    
\end{enumerate}

The adopted measures are meant to characterize the network locally and globally. 


Because the network features can be extracted along the evolution of the network reconstruction, two strategies for the network characterization were considered. The strategy referred to as ``last value'' considered only the network generated at the end of the random walk.  We also considered the approach referred to as ``all values''. Here, the above measurements are extracted considering the evolution of the network as symbols are generated and incorporated in the network. For each random walk length, we extracted the features along the evolution by considering 10 intermediary values. For example, for the random walk considering 1,000 steps, we extracted the network features when the agent completes $100$, $200$, $300$  $\ldots$ steps. 



\subsection{Topology and dynamics classification} \label{sec:classification}

The properties extracted from the reconstructed  network are features that are used to characterize samples in multi-class classifiers~\cite{amancio2014systematic}, where classes can be: (i) the topology of the network; (ii) the dynamics used by the agent; and (iii) both (i) and (ii). This experiment aims to understand how accurate one can estimate the topology and dynamics generating the observed sequence of symbols.


The goal of the classification, as discussed previously, is to obtain a classifier model that can successfully classify the samples, this case the properties of the generated network, into known structures. Therefore, we may have a generalist model that will be able to predict unknown sequences and infer which topology and/or dynamics generated such sequence, based only on the resulting properties of the sequence's reconstructed network.

In order to evaluate whether the extracted properties can be used to detect the mechanisms and structure generating a sequence, we used well-known supervised classifiers~\cite{amancio2014systematic}. 
%
The following classifiers  were chosen to detect patterns in the data: Decision Trees (DT)~\cite{Quinlan1986decisiontree}, Random Forest~(RF)
\cite{Breiman2001randomforest}, Stochastic Gradient Descent (SGD)~\cite{Bottou1998SGD}, Multilayer Perceptron (MLP)~\cite{Haykin1998MLP}, k-Nearest Neighbors (KNN)~\cite{Fix1989KNN}, Linear Discriminant Analysis (LDA) \cite{hastie2009LDA}, and Gaussian Naive Bayes (GNB) \cite{Raizada2013GNB}. The parameters were optimized according to the approach suggested in related works~\cite{amancio2014systematic}.




\section{Results and discussion} \label{sec:results}


\subsection{Evolution of the reconstructed network}

\begin{figure}
\centering\includegraphics[width=1.0\linewidth]{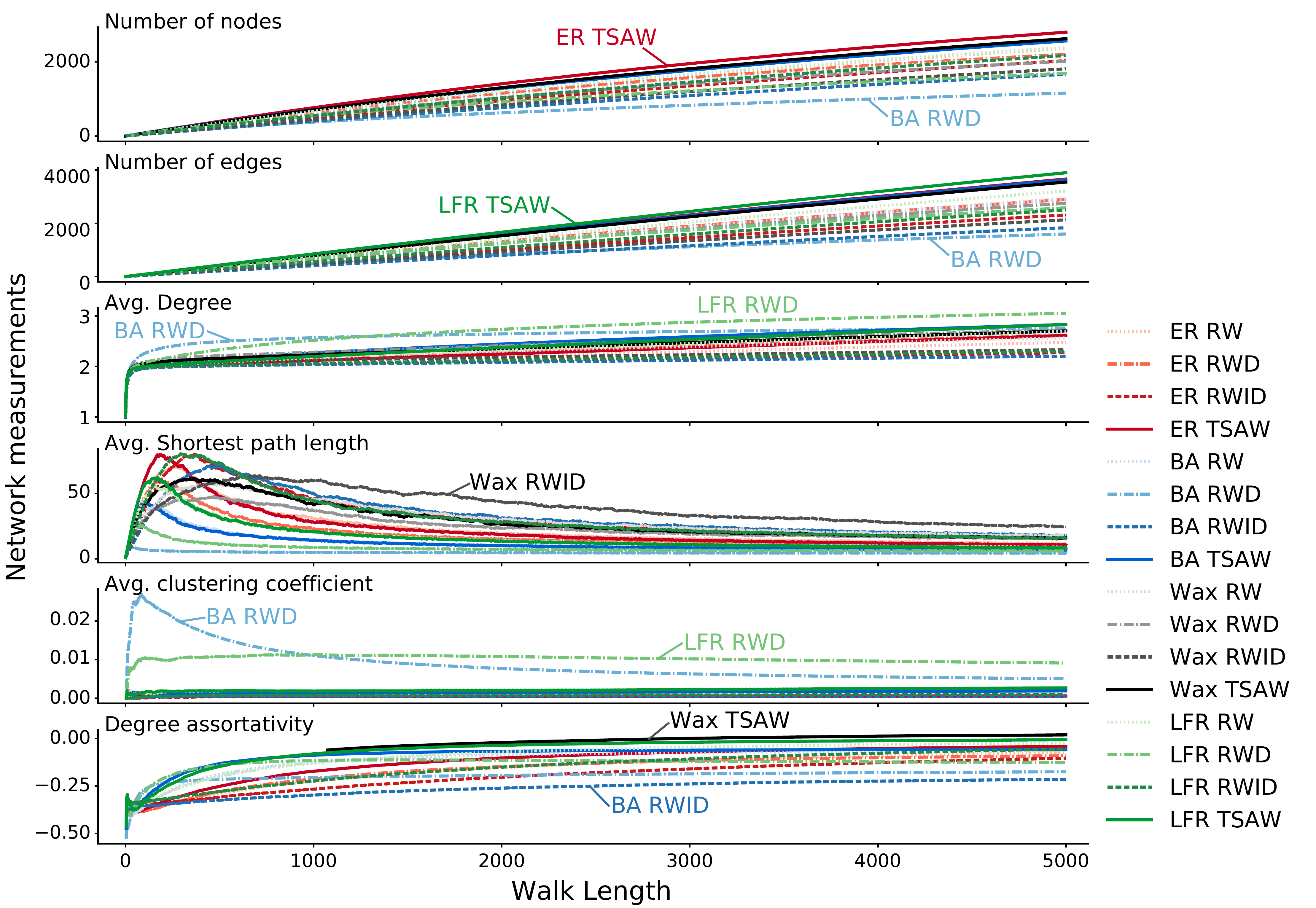}
\caption{Evolution of the network metrics obtained from the reconstructed networks when considering different values of walk length. The original network was created considering $\langle k\rangle = 4$ in all network models.}.
\label{fig:propertiesk4}
\end{figure}

To analyze the evolution of the considered metrics as network are reconstructed, we show, in Figure \ref{fig:propertiesk4} the values of the considered metrics for different walk lengths, models and dynamics. Here we considered  $\langle k\rangle = 4$, however similar results were obtained for different values of average degree (result not shown). 


The results show a similar behavior for LFR and BA models with the RWD dynamics in most properties curves. Also, we observe that the curves for both LFR and BA when considering the RWD dynamics correspond to the smallest growth along the walk with a large gap between them and the other dynamics that increases as more steps are considered (i.e. as the network grows). The results also revealed that is that the TSAW dynamics displayed the highest efficient to discover nodes and edges. Concerning the discovery of edges, we can notice that the gap between TSAW and other dynamics is even larger. These results achieved by the TSAW dynamics are consistent with similar findings~\cite{Guerreiro2020}. 

As for the average degree, we found that the RWD dynamics on both LFR and BA models discovers new edges much faster than nodes, leading thus to a higher estimated degree compared to the other combination of models an dynamics. This effect might stem from the fact that the TSAW takes into account the visited edges, so already visited edges tend to be less visited in future interactions. %




Interestingly, the shortest path length initially takes high values in almost all cases. This happens because when new nodes are discovered and not revisited, the network is similar to a line graph.  For larger values of walk length, the shortest path length seems to converge. Once again the RWD dynamics for both LFR and BA networks displayed a different behavior: the estimated value along the first steps are not as high as the ones observed for the other combinations of network and dynamics. 

For the clustering coefficient property, while most curves remain close to each other, LFR and BA for the RWD dynamics, again, seem to estimate much higher values of cluster coefficient compared to the other configurations. This might be a consequence of this dynamics being sensitive to hubs, which is present in both LFR and BA topologies. 
Finally, for the degree assortativity, after larger variation for short sequences, the assortativity seems to increase (and converge) for all combinations of network models and dynamics as more nodes are visited along the walk. 

In general, simpler features, such as the number of nodes, edges and the average degree and degree associativity resulted in monotonically increasing curves, with no crosses or substantial changes between. In contrast, different patterns where observed for the clustering coefficient and shortest path length. 

\subsection{Detecting the network topology and walk dynamics}

While in Figure \ref{fig:propertiesk4} one can observe in fact some differences in the behavior of curves, the discrimination between network models, dynamics or even both is not visually evident. Because we are interested in recovering the original network model and dynamics that generated the sequence (i.e. the first two steps in Figure \ref{fig:schematic}), we conducted a machine learning experiment. Our machine learning experiment was carried out in a threefold way, depending on the information being recovered: network model, agent dynamics, or both. For each configuration of network model, we considered 100 random realizations. Concerning the different dynamics, each realization started from a random seed, and 1,000 realizations were considered. 


In Table \ref{tab:model} we show the accuracy rate obtained when identifying the model generating the sequence. Two approaches to compose the features are considered. The approach referred to as ``all'' extract the features obtained as the reconstructed network evolves, while ``last'' only considers the last configuration of the reconstructed network. We show the results obtained for different walk lengths $s = \{10,50,100,500,1000,2000,5000\}$. The results show that -- as expected -- the discriminability is not significant for short walks. However, it is interesting to note that an accuracy higher than 50\% can be obtained with only 100 steps (for the RF classifier). The best accuracy also significant is improved when considering 500 steps, reaching an accuracy higher than 86\% when less than 20\% of network is discovered. An almost perfect discrimination of network models is achieved with a sequence of 5000 symbols (walk length). It is also worth noting also that, in almost all cases, the best classification is obtained with the Random Forest classifier. In almost all scenarios, we also noted that the performance considering the ``last'' and ``all'' approaches are similar, meaning that the evolution of the reconstructed network does not provide a significant amount of information for this classification task. 

\begin{table}[h]
\centering
\caption{Accuracy rate (\%) for the classification considering the four network models: ER, BA, Wax and LFR (see Section \ref{sec:topologies}). Columns represent the number of steps taken by the agents. When considering 5,000 steps in the walk, the best result is obtained with the Random Forest classifier. The network model generating the sequence can be identified with an accuracy of 98.59\%. The best results for each walk length are highlighted.}
\begin{tabular}{| l | c |  c | c | c | c | c | c | c | }
\hline
  \multicolumn{2}{|c|}{{\bf Method}}       &    \ \ 10	\ \ 	&  \ \	50 \ \		&      \    100 \      &     \      500 \        &     1,000        &          2,000         &        5,000   \\
\hline
\multirow{2}{*}{DT}	& last         &     28.7        &     42.0       &     47.1       &    69.9        &    81.6        &    90.5        &     97.6     \\
				& all           &     28.9        &     35.4       &     43.3       &    70.6        &    81.6        &    90.7       &      97.8     \\
\hline
\multirow{2}{*}{RF}	& last         &     28.7       &      42.9      &      49.0      &     75.1       &     85.6       &     93.0      &       98.3     \\
				& all           &      \textbf{29.0}       &    42.0        &      \textbf{53.0}      &     \textbf{78.1}       &      \textbf{86.6}       &     \textbf{93.5}      &       \textbf{98.6}      \\
\hline
\multirow{2}{*}{SGD}	& last         &     25.8       &     29.7       &     36.7       &      49.0      &     48.4       &    47.6       &     49.4       \\
				& all           &     25.7       &     28.2        &    40.6       &      45.0      &     53.5       &    43.7       &     52.8       \\
\hline
\multirow{2}{*}{MLP}	& last         &     28.2       &     43.3       &    52.1       &       72.5     &      80.0      &    86.2       &      76.3      \\
				& all           &     28.3       &     \textbf{43.6}       &    52.3      &      74.5      &      80.8      &    80.5       &    39.1       \\
\hline
\multirow{2}{*}{KNN}	& last         &     26.5       &       37.3     &      46.4     &     65.5      &     72.4       &      80.4     &      91.0      \\
				& all           &     26.1       &       34.0    &      40.1     &     63.8      &     74.2       &      82.2     &      93.6      \\
\hline
\multirow{2}{*}{LDA}	& last         &     28.0       &      37.6      &     41.9       &    58.4       &     64.4      &     72.5      &      74.0      \\
				& all           &     28.0       &      38.2      &     42.8       &     60.6      &     67.7      &     73.8      &      81.5      \\
\hline
\multirow{2}{*}{GNB}	& last         &     27.2       &      37.7      &     45.1       &     53.7       &     55.7       &    57.9       &    53.8        \\
				& all           &     26.5       &      35.5      &     42.0       &     53.5       &     55.3       &    56.3      &     58.7       \\
\hline
\end{tabular}
\label{tab:model}
\end{table}

In Table \ref{tab:dynamics} we show the accuracy rates obtained when detecting the walk dynamics used to generate the sequence. The results show that the best accuracy rate is similar to the one obtained when identifying the network models ($98.76\%$). 
The dynamics are retrieved with higher accuracy than models for short sequences (10, 50 and 100), while model recovery accuracy is higher for longer sequences (500, 1000, and 2000). Surprisingly, when less than 2\% of the network is recovered (100 steps) one is able to identify the walk dynamics with an accuracy higher than 50\%. 
Likewise, one can reach almost 80\% of accuracy when less than 10\% of the network is discovered (500 nodes). 
In order to recover the dynamics with an accuracy higher than 85\%, 1000 steps are required. 
Concerning the methods, the Random Forest classifier once again achieved most of the highest accuracies, along with the ``all'' strategy. The MLP algorithm displayed competitive results, specially for the sequence sizes of 50 and 100. 

\begin{table}[h]
\centering
\caption{Accuracy rate (\%) for the classification considering the four network dynamics: RW, RWD, RWID and TSAW (see Section \ref{sec:topologies}).  Columns represent the number of steps taken by the agents. When considering 5,000 steps in the walk, the best result is obtained with the Random Forest classifier. The network dynamics generating the sequence can be identified with an accuracy of 98.76\%. The best results for each walk length are highlighted.}
\begin{tabular}{| l | c |  c | c | c | c | c | c | c | }
\hline
  \multicolumn{2}{|c|}{{\bf Method}}       &    \ \ 10	\ \ 	&  \ \	50 \ \		&      \    100 \      &     \      500 \        &     1,000        &          2,000         &        5,000   \\
\hline
\multirow{2}{*}{DT}	& last         &      40.0       &       54.3     &      55.3      &      65.9      &     77.3       &      89.1      &     97.9     \\
				& all           &      \textbf{40.4}       &     46.9       &      51.3      &     65.8       &      76.6      &     88.7      &    97.9       \\
\hline
\multirow{2}{*}{RF}	& last         &      40.0      &      54.7      &     59.9       &     71.0       &     81.8       &    92.0       &     98.8       \\
				& all           &     \textbf{40.4}       &     54.2       &     61.1       &      \textbf{72.9}       &     \textbf{82.2}       &      \textbf{92.1}     &     \textbf{98.8}        \\
\hline
\multirow{2}{*}{SGD}	& last         &      37.8      &     51.7       &     56.4       &     62.8       &     63.7       &     57.3      &     66.0      \\
				& all           &      38.2      &     51.8        &    57.3       &     63.8       &     63.3       &     69.7      &     70.6       \\
\hline
\multirow{2}{*}{MLP}	& last         &     39.7       &     55.8       &      60.7     &     67.0       &     71.3       &     78.7      &     83.7       \\
				& all           &     40.3       &     \textbf{56.0}       &     \textbf{61.4}     &     69.0       &      72.3      &       75.8    &      87.6      \\
\hline
\multirow{2}{*}{KNN}	& last         &    27.7        &     50.9       &     56.1      &     66.8     &    73.5        &   82.4        &      92.5      \\
				& all           &    29.3        &      49.7      &      55.5     &    65.0       &   72.4         &   83.4        &    95.1        \\
\hline
\multirow{2}{*}{LDA}	& last         &    39.7        &     53.8       &    57.9        &     62.0      &     64.4      &    66.7       &    69.3        \\
				& all           &    39.6        &     54.2       &    58.2        &     63.2      &     65.2      &    68.1       &   79.2        \\
\hline
\multirow{2}{*}{GNB}	& last         &    36.0        &     51.8       &    55.8        &    61.6        &   63.1         &   63.1        &    60.6        \\
				& all           &     35.6       &      49.4      &    54.2        &    60.6        &    62.4        &    63.4      &     62.6       \\
\hline
\end{tabular}
\label{tab:dynamics}
\end{table}

Table \ref{tab:modelanddynamics} shows the results obtained when identifying both the network topology and walk dynamics generating the observed sequence. While this is a much difficult task since we are discriminating 16 classes (i.e. four different topologies and four different walks), high accuracy rates can be obtained, specially when for walks comprising more than 1000 steps. At the best scenario, the accuracy rate reaches $97.57\%$. When very short walks are considered, the generated sequence does not provide much discriminative information, which leads to typical low accuracy rates. The results suggest that it is possible to retrieve the originating model and dynamics by looking only to the reconstructed network properties, given that the sequence length is long enough. 


\begin{table}[h]
\centering
\caption{Accuracy rate (\%) for the classification considering both network models and walk dynamics: \{ER, BA, Wax, LFR\} $\times$ \{RW, RWD, RWID, TSAW\} (see Section \ref{sec:topologies}).  Columns represent the number of steps taken by the agents. When considering 5,000 steps in the walk, the best result is obtained with decision trees from the Random Forest classifier. Both the network topology and  walk dynamics generating the sequence can be identified with an accuracy of 97.57\%. The best results for each walk length are highlighted.}
\begin{tabular}{| l | c |  c | c | c | c | c | c | c | }
\hline
  \multicolumn{2}{|c|}{{\bf Method}}       &    \ \ 10	\ \ 	&  \ \	50 \ \		&      \    100 \      &     \      500 \        &     1,000        &          2,000         &        5,000   \\
\hline
\multirow{2}{*}{DT}	& last         &       12.1       &       22.3      &       27.0      &      49.3       &      65.6       &      82.1       &      96.0       \\
				& all           &       \textbf{12.4}       &       18.2      &       24.5      &      49.9       &      65.4       &      81.9      &      96.3      \\
\hline
\multirow{2}{*}{RF}	& last         &       12.1      &       23.0      &        28.8     &        55.3     &       72.1      &       86.7     &       97.2      \\
				& all           &       \textbf{12.4}      &     23.2         &     32.2        &     \textbf{59.2}        &        \textbf{73.4}     &     \textbf{87.3}       &      \textbf{97.6}       \\
\hline
\multirow{2}{*}{SGD}	& last         &      7.9       &     16.4        &     19.6        &      32.3       &     30.7        &      35.8      &       29.2      \\
				& all           &      7.1      &     12.8         &    23.5         &      28.8       &     39.2        &      40.3      &      42.8       \\
\hline
\multirow{2}{*}{MLP}	& last         &      11.7       &       24.3      &      31.7       &      54.4       &     65.1        &      75.6      &       86.0      \\
				& all           &     12.1        &       \textbf{25.0}       &     \textbf{33.0}        &      55.1       &       63.4      &      74.3      &       88.4       \\
\hline
\multirow{2}{*}{KNN}	& last         &     8.3        &      19.6       &      26.7       &       46.0      &      57.4       &      71.7      &     87.8        \\
				& all           &     10.0       &       17.9      &       23.5      &        42.6     &       56.5      &       72.8     &      91.0       \\
\hline
\multirow{2}{*}{LDA}	& last         &     11.7        &      21.4       &      26.4       &       41.9      &      49.1       &      55.4      &      66.2       \\
				& all           &     11.9        &      21.9       &      27.2       &       43.0      &      50.3       &      58.6      &      74.8       \\
\hline
\multirow{2}{*}{GNB}	& last         &     10.0        &      21.1       &     27.1        &      43.6       &      47.8       &    48.8       &     49.2        \\
				& all           &    9.7         &      18.9       &     25.0        &      41.4       &      46.7       &     49.1      &      50.8       \\
\hline
\end{tabular}
\label{tab:modelanddynamics}
\end{table}

\begin{figure}[h]
\centering\includegraphics[width=0.8\linewidth]{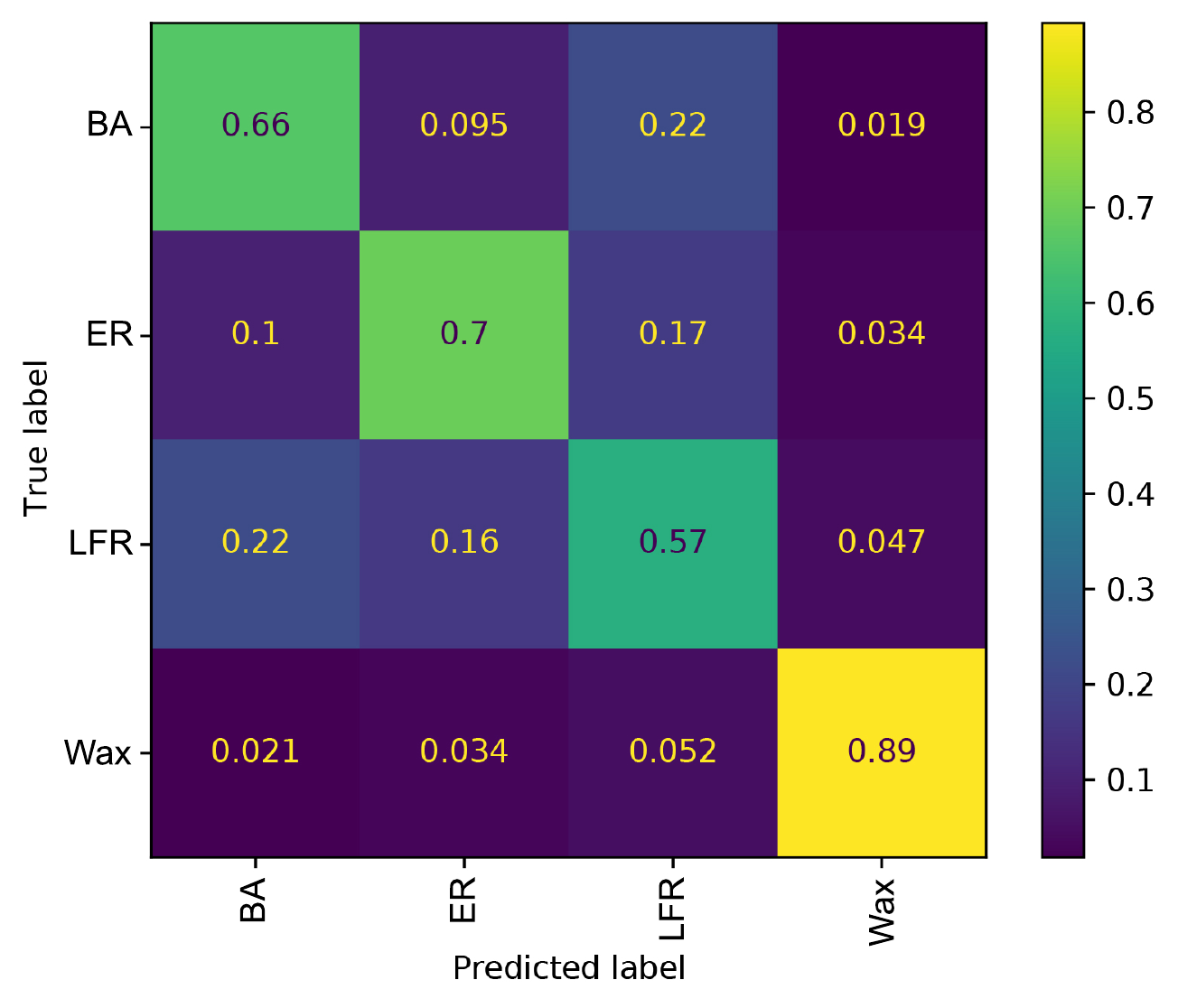}
\caption{Confusion matrix considering decision trees in the classification of sequences. The classification considered four network models: Erd\H{o}s-Rényi (ER), Barabási-Albert (BA), Waxman (Wax) and Modular Networks (LFR). }
\label{fig:confmodels}
\end{figure}

\begin{figure}[h]
\centering\includegraphics[width=0.8\linewidth]{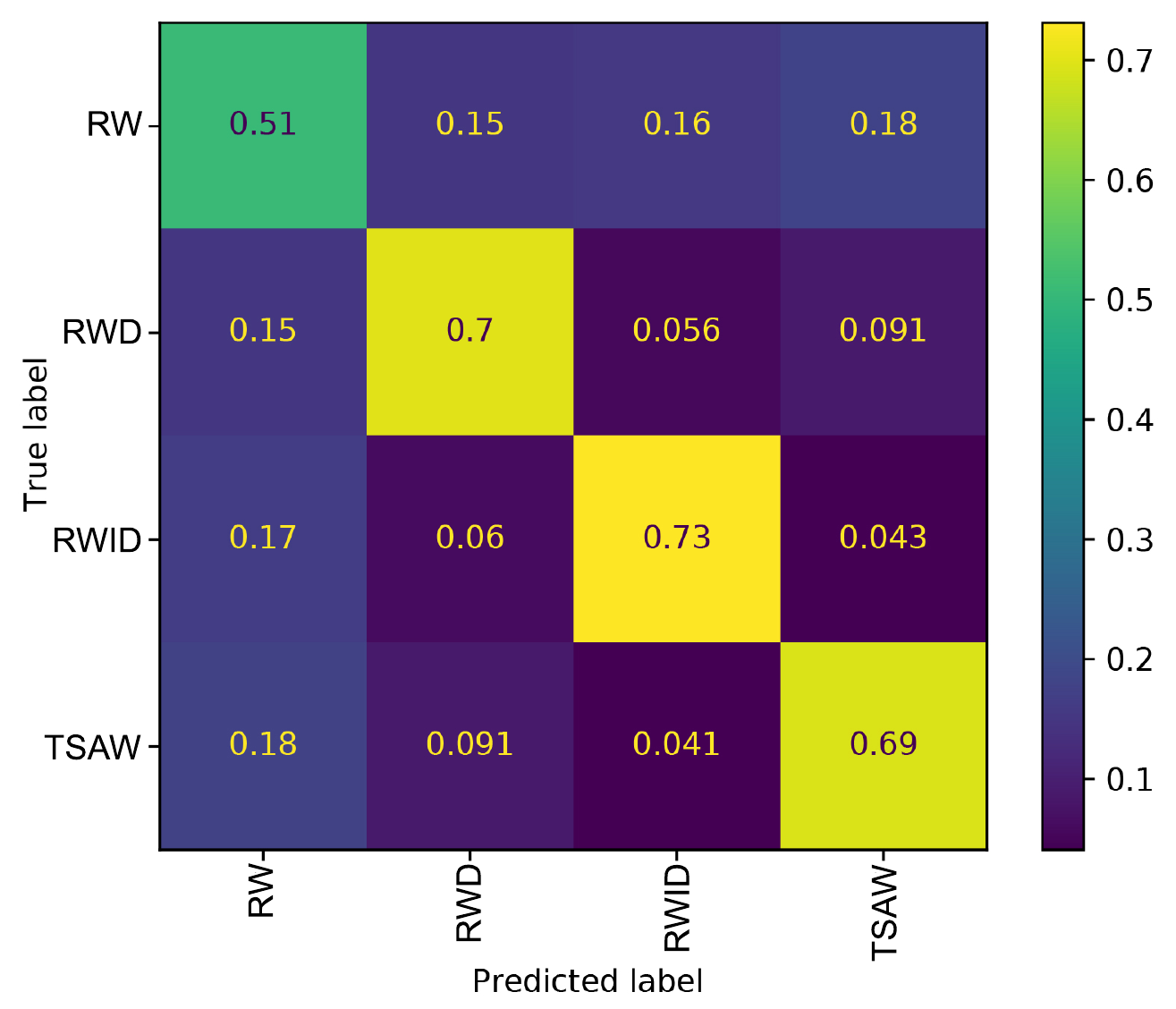}
\caption{Confusion matrix considering decision trees in the classification of sequences. The classification considered four network dynamics: random walk (RW), degree-biased random walk (RWD), inverse degree biased random walk (RWID) and true self-avoiding random walk (TSAW).}
\label{fig:confdynamics}
\end{figure}

In order to better understand the errors in the adopted classifiers, we analyzed confusion matrices. We selected the matrix corresponding to the results obtained with 500 steps because a higher number of steps usually leads to a very high accuracy rates. Conversely, very short walks usually leads to higher error rates mainly because almost all of the network can not be discovered with only a few steps. Similar results were also obtained when considering 1000 steps. 
The confusion matrix obtained from the classification of network models and dynamics are displayed in Figures \ref{fig:confmodels} and \ref{fig:confdynamics}, respectively. 


The confusion matrix obtained for the network models reveals that the Wax model can be predicted with the highest accuracy. If a sequence is generated from a Wax network, it can be predicted with an accuracy of roughly $89\%$. Conversely, network with community structure are the ones generating the highest error rates. 22\% of all sequences generated on LFR networks are classified as a BA network. While this can be explained by the fact that both BA and LFR networks present long tail distributions, they still have different mesoscale organization.  A high error rate also associates LFR sequences as if they were generated on ER networks.




The confusion matrix for the agent dynamics classification (Figure \ref{fig:confdynamics}) shows that RWID is the one with the highest accuracy: if the sequence is generated via a RWID walk, it can be recovered with an accuracy of 73\%.
In a similar fashion, RWD and TSAW had similar results in terms of general accuracy. Most errors in the RWID classification occurred as RW predictions, and the same behavior can be observed for RWD and TSAW, i.e., most of the mispredictions in this classifier occurred as RW guesses. Interestingly, random walks has a low accuracy. Its behavior is classified as RWD, RWID and TSAW dynamics with similar probability. In sum, the lack of accuracy in the classification of dynamics occur mostly when detecting that walk is the traditional unbiased random walk. This lack of correspondence might happen when the agent explores homogeneous region, causing thus no distinctive effect for degree-biased random walk. In a similar fashion, when the network is not fully explored, the true self-avoiding walk is seldomly applied and the TASW behaves like a RW walk.



\begin{figure}[h]
\centering\includegraphics[width=1\linewidth]{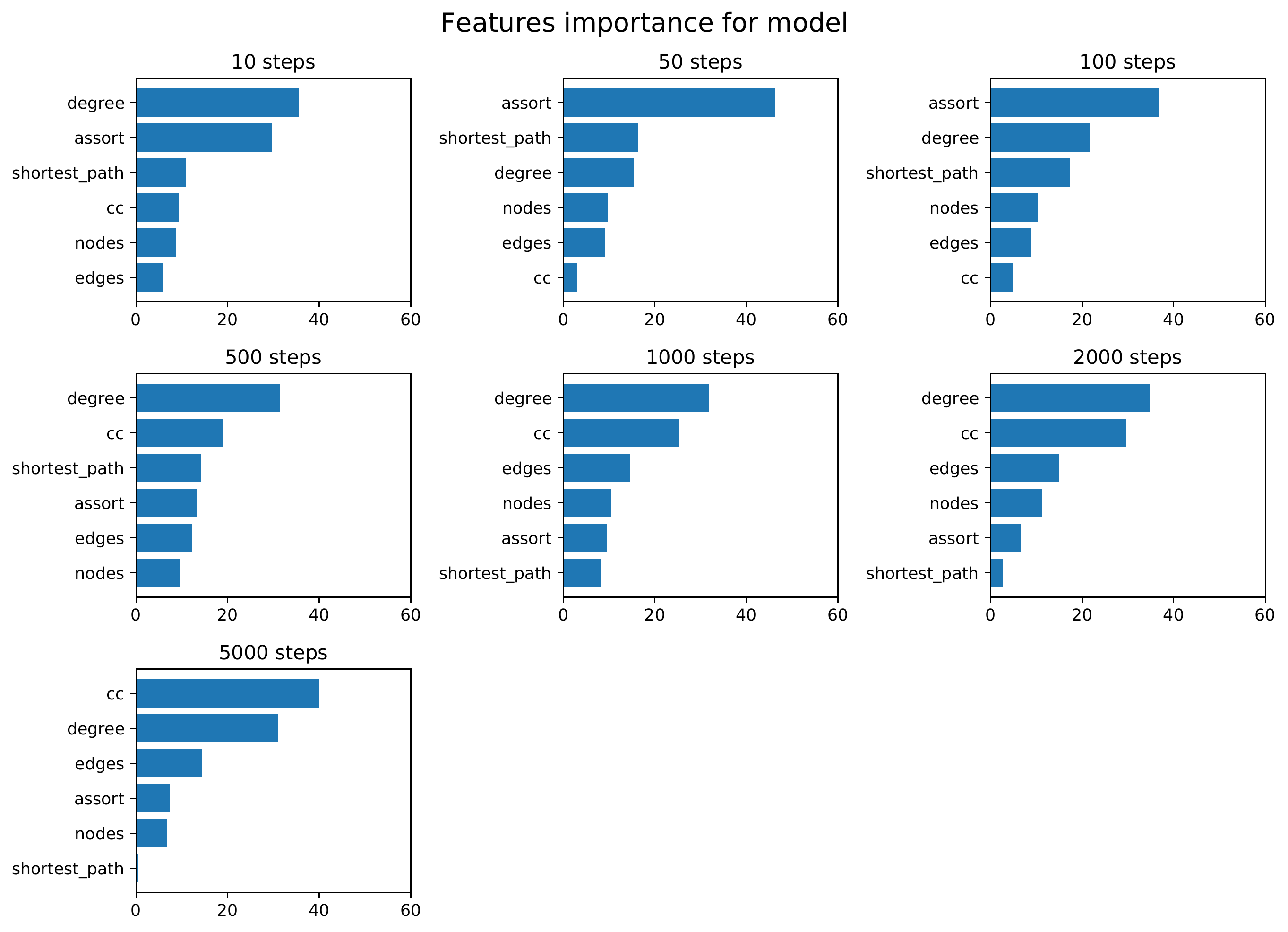}
\caption{Features importance for the classification of \emph{network models}. The values in the x-axis are proportional to the features relevance in the classification task. }
\label{fig:featmodel}
\end{figure}
\begin{figure}[h]
\centering\includegraphics[width=1\linewidth]{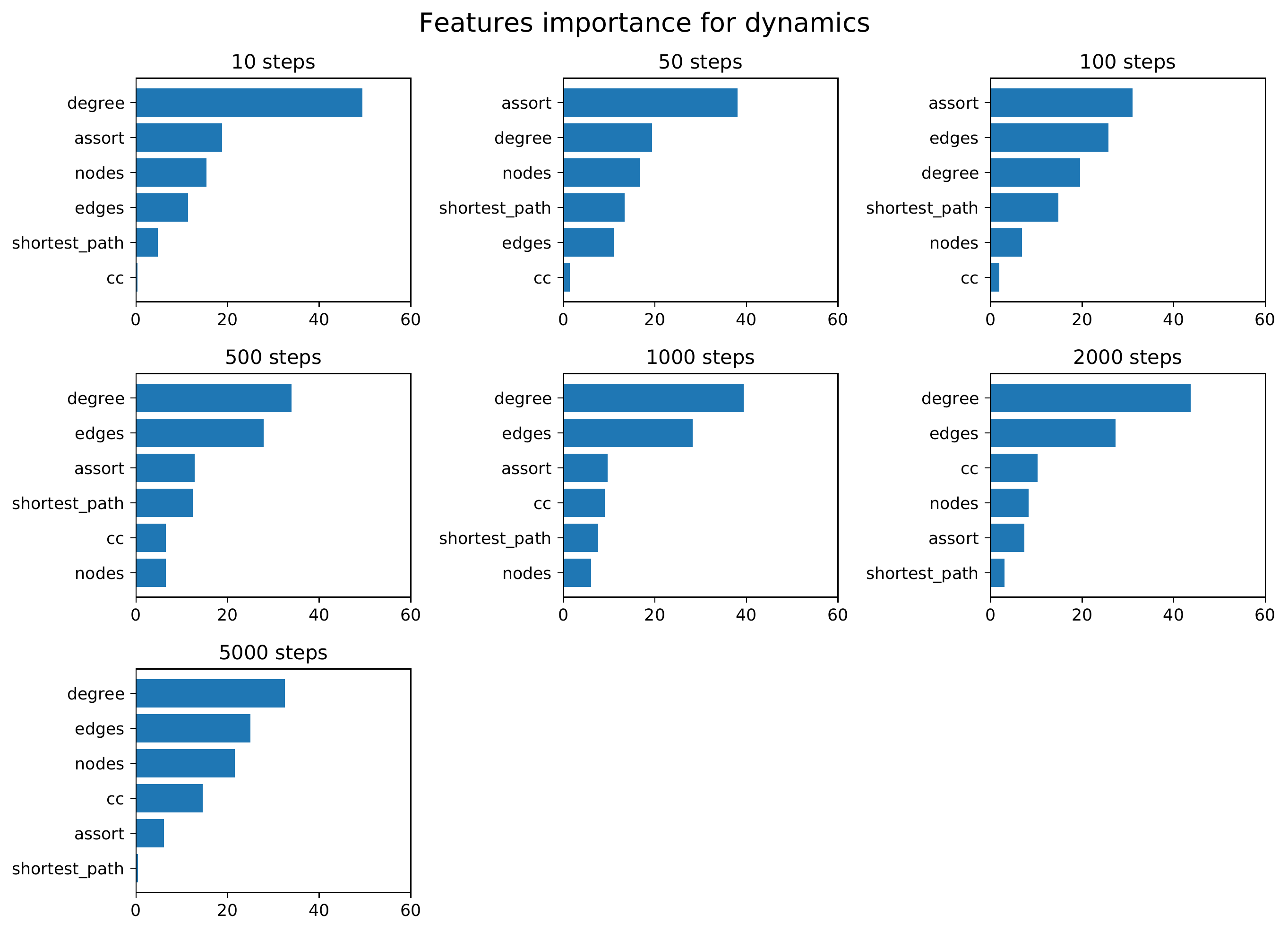}
\caption{Features importance for the classification of \emph{network dynamics}. The values in the x-axis are proportional to the features relevance in the classification task.}
\label{fig:featdynamics}
\end{figure}

In order to better understand the contribution of the adopted features (see Section \ref{sec:properties}) to recover the structure and dynamics generating the observed sequences, we probed features are the most relevant in the classification task. We used the Gini relevance index  to compute the relevance of the features. This index is used in many different contexts~\cite{nembrini2018revival,tohalino2022predicting}. The most relevant features for all the sequences sizes studied for model and dynamics are shown in Figures \ref{fig:featmodel} and \ref{fig:featdynamics}, respectively. For each subplot, the features are sorted in decreasing order of importance. We only show the results for the ``all'' approach.

%

We found a similar behavior for all three classification scenarios. The degree property is a relevant feature for most of the sequence sizes, meaning that the relationship between vocabulary size (i.e. the number of different nodes discovered~\cite{Guerreiro2020}) and the sequence length is a discriminative feature. 
Conversely, the shortest path property is typically the least relevant feature in larger sequences, while being among the most relevant features for smaller sequences.
The total number of edges is not a relevant feature for smaller sequences, but it is one of the most important for larger sequences; therefore, as the network structure has more nodes, the number of edges becomes a significant feature.  All in all, the results show that the best accuracy is obtained with larger sequences, and in those scenarios, local features such as the number of nodes, edges and clustering coefficient are among the most relevant features to identify the network model and agent dynamics generating the observed sequences.

\section{Conclusion}

In the current paper, we analyzed whether the observation of sequences as a result of agent walking over a network topology can provide accurate information regarding the process generating the sequence. We used the observed sequence to reconstruct a network via co-occurrence links, and then the observed network properties were used to infer both the original network models and the rule used by the agent to walk on the network. The properties of the reconstructed network were then used to characterize the reconstructed network in machine learning models. Our experiments were performed using 4 topological models and 4 random walk rules. 

Our results revealed that one can predict both the network topology and agent dynamics with high accuracy provided that the observed sequence has a minimum length. When predicting only the network topology, the correct topology could be found with accuracy higher than 86\% when less than 20\% of nodes are visited. This accuracy increases to 93\% when visiting less than 40\% of nodes. When predicting the walk rule used by the agent, we found a slight lower accuracy. When 20\% and 40\% of all nodes were visited, we recovered the agent dynamics with an accuracy of 82\% and 92\%, respectively. Our models were are trained to infer both the topology and agent dynamics in the same model. In this case, the model also displayed excellent performance, especially for walk length larger than 40\% of the network. When considering shorter sequences, we found that distinct combinations of network models and walk dynamics result in different performances.

We showed, as a proof of principle, that it is possible to recover the generating model and dynamics from features extracted from reconstructed networks. Future works could address the problem of finding the best method to reliably recover the most suitable network structure and dynamics from sequences. Another question is checking how robust the measures are across different strategies and sizes to reconstruct networks from sequences.
In terms of applications, the developments of this work could be applied to identify patterns  in real-world sequences, such as clickstreams~\cite{aguiar2016digital} across the web, or from users exploring social media and video content. In such case, the dynamics performed sequences and those suggested by recommendation systems could be compared.






 
\section*{Acknowledgments}

D.R.A. acknowledges financial support from CNPq-Brazil (grant no. 311074/2021-9) and São Paulo Research Foundation (FAPESP grant no. 20/06271-0).

\newpage

\bibliographystyle{ieeetr}
\bibliographystyle{abbrv}

\end{document}